\documentclass[12pt]{iopart}

\usepackage[plainpages=false,breaklinks=true]{hyperref}
\usepackage{graphicx,color}
\usepackage{tabularx}
\usepackage{iopams}

\usepackage{framed}

\usepackage{float}

\newcommand{\calv}{C_{\mathrm{A}}}

\newcommand{\cart}{C_{\mathrm{a}}}
\newcommand{\cven}{C_{\bar{\mathrm{v}}}}
\newcommand{\cinh}{C_{\mathrm{I}}}

\newcommand{\cper}{C_{\mathrm{per}}}
\newcommand{\crpt}{C_{\mathrm{rpt}}}
\newcommand{\cbody}{C_{\mathrm{B}}}

\newcommand{\lrpt}{\lambda_{\mathrm{b:rpt}}}
\newcommand{\lper}{\lambda_{\mathrm{b:per}}}
\newcommand{\hen}{\lambda_{\mathrm{b:air}}}

\newcommand{\qalv}{\dot{V}_{\mathrm{A}}}

\newcommand{\qc}{\dot{Q}_{\mathrm{c}}}

\newcommand{\prl}{k_\mathrm{pr}^{\mathrm{rpt}}}
\newcommand{\prm}{k_\mathrm{pr}^{\mathrm{per}}}
\newcommand{\ml}{k_\mathrm{met}^{\mathrm{rpt}}}
\newcommand{\mm}{k_\mathrm{met}^{\mathrm{per}}}
\newcommand{\mbody}{k_\mathrm{met}} 
\newcommand{\prbody}{k_\mathrm{prod}}

\newcommand{\valv}{\tilde{V}_{\mathrm{A}}}

\begin{document}

\title[Modeling-based determination of physiological parameters of systemic VOCs]{Modeling-based determination of physiological parameters of systemic VOCs by breath gas analysis: a pilot study}

 \author{Karl Unterkofler$^{1,2}$,
 	Julian King$^1$,
 	Pawel Mochalski$^1$,
 	Martin Jandacka$^{1,2}$,
 	Helin Koc$^1$,
 	Susanne Teschl$^3$,
 	Anton Amann$^{1,4}$, and
 	Gerald Teschl$^5$ }
\address{$^1$Breath Research Institute, University of Innsbruck, Rathausplatz 4, A-6850 Dornbirn, Austria}
\address{$^{2}$University of Applied Sciences Vorarlberg, Hochschulstr.\ 1, A-6850 Dornbirn, Austria}
\address{$^{3}$University of Applied Sciences Technikum Wien,  H\"ochst\"adtplatz 6, A-1200 Wien, Austria}
\address{$^4$Univ.-Clinic for Anesthesia and Intensive Care, Innsbruck Medical University, Anichstr. 35, A-6020 Innsbruck, Austria}
\address{$^{5}$Faculty of Mathematics, University of Vienna, Oskar-Morgenstern-Platz 1, 1090 Wien, Austria}
  
\ead{gerald.teschl@univie.ac.at}
\ead{karl.unterkofler@fhv.at}

\begin{abstract}
In this paper we develop a simple two compartment model which extends the Farhi 
equation to the case when the inhaled concentration of a volatile organic compound (VOC) is not zero. 
The model connects the exhaled breath concentration of systemic VOCs with physiological 
parameters such as endogenous production rates and metabolic rates. Its validity
is tested with data obtained for isoprene and inhaled  deuterated isoprene-D5.
\end{abstract}
Version: \today\\
J. Breath Res. {\bf 9}, 036002 (2015).

\maketitle

\section{Introduction} 
 
The advance of different analytical methods in mass spectrometry within the last twenty years
has opened the door to breath gas analysis. 
There is considerable evidence that volatile organic compounds
(VOCs) produced in the human body and then partially released in breath 
 have great potential for diagnosis in physiology and
medicine \cite{Costello2014a}. The emission of such compounds may result
from normal human metabolism as well as from pathophysiological
disorders, bacterial or mycotic processes (see \cite{amannbook2013} and the references therein), or
exposure to environmental contaminants \cite{Pleil2013, Pleil2013b, wallace1993}. As subject-specific
chemical fingerprints, VOCs can provide non-invasive
and real-time information on infections, metabolic disorders,
and the progression of therapeutic intervention.

In a recent paper Spanel et al.\ \cite{spanel2013} investigated the short-term effect of inhaled VOCs on their exhaled breath concentrations.
They showed for seven different VOCs that the exhaled breath concentration closely resembles an affine function of the inhaled concentration.
This  motivated our theoretical investigation regarding the impact of inhaled concentrations for VOCs with low blood:air partition coefficients, i.e., compounds with exhalation kinetics that are described by the Farhi equation \cite{Farhi1967e}. 

 To this extent we develop a simple two compartment model which generalizes the Farhi 
 equation to the case in which the inhaled concentration of a VOC is not negligible. 
In accordance with the above-mentioned experimental observations, the model predicts that when ventilation and perfusion are kept constant the exhaled breath concentration is indeed an affine function of the inhaled concentration.
In addition it links the exhaled breath concentration of systemic VOCs to physiological parameters such as endogenous production rates and metabolic rates, thereby complementing similar efforts in the framework of exposure studies \cite{pleilbook, csanady2001}. This estimation process is exemplified by means of 
exhalation data for endogenous isoprene and inhaled deuterated isoprene-D5.

Another interesting aspect of the model is that   for low-soluble VOCs it illustrates a novel approach for answering the question ``Is subtracting the inhaled concentration from the exhaled concentration a suitable method to correct measured breath concentrations for room air concentrations?'', an issue that is still being debated within the breath analysis community \cite{phillips1999, schubert2005}.
   In the discussions we indicate how to extend these results to VOCs with higher partition coefficients and how to take into account long-term exposure.\\
   A list of symbols used is  provided in  Appendix~A.
  
  \section{A two compartment model} \label{section2}

\subsection{Derivation of  the Farhi equation} \label{farhi}

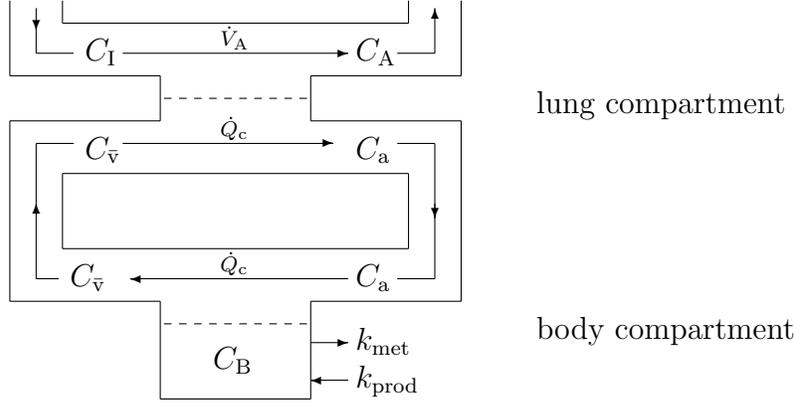
\begin{figure}[ht]
\centering
\centering
\begin{picture}(11.5,6.5)

\put(8,5.3){lung compartment}
\put(8,2.3){body compartment}

\put(1,5.8){\line(0,1){1}}
\put(1.7,6.5){\line(0,1){0.3}}
\put(7,5.8){\line(0,1){1}}
\put(6.3,6.5){\line(0,1){0.3}}

\put(1.7,6.5){\line(1,0){4.6}}
\put(1,5.8){\line(1,0){2}}\put(5,5.8){\line(1,0){2}}
\put(3,5.8){\line(0,-1){0.6}}
\put(5,5.8){\line(0,-1){0.6}}
\put(1,5.2){\line(1,0){2}}\put(5,5.2){\line(1,0){2}}
\put(1.7,4.5){\line(1,0){4.6}}

\multiput(3.05,5.5)(0.2,0){10}{\line(1,0){0.1}}

\put(2,6){$\cinh$}
\put(2,4.7){$\cven$}

\put(5.6,6){$\calv$}
\put(5.6,4.7){$\cart$}

\put(2.5,6.1){\vector(1,0){3}}
\put(3.8,6.2){$\scriptstyle\qalv$}
\put(1.35,6.1){\line(1,0){0.5}}
\put(1.35,6.7){\vector(0,-1){0.3}}
\put(1.35,6.4){\line(0,-1){0.3}}
\put(6.15,6.1){\line(1,0){0.5}}
\put(6.65,6.1){\vector(0,1){0.6}}

\put(2.5,4.9){\vector(1,0){2.8}}
\put(3.8,5){$\scriptstyle\qc$}

\put(1,5.2){\line(0,-1){2.4}}
\put(7,5.2){\line(0,-1){2.4}}

\put(1.7,4.5){\line(0,-1){1}}
\put(6.3,4.5){\line(0,-1){1}}

\put(1.7,3.5){\line(1,0){4.6}}
\put(1.0,2.8){\line(1,0){2.0}}
\put(5,2.8){\line(1,0){2.0}}
\put(3,2.8){\line(0,-1){1.3}}
\put(5,2.8){\line(0,-1){1.3}}
\put(3,1.5){\line(1,0){2}}

\multiput(3.05,2.5)(0.2,0){10}{\line(1,0){0.1}}

\put(5.5,3.1){\vector(-1,0){2.9}}
\put(3.8,3.2){$\scriptstyle \qc$}

\put(5.6,3){$\cart$}
\put(1.8,3){$\cven$}
\put(3.7,1.9){$\cbody$}
\put(5,2.25){\vector(1,0){0.5}}
\put(5.6,2.15){$\mbody$}
\put(5.5,1.75){\vector(-1,0){0.5}}
\put(5.6,1.65){$\prbody$}

\put(1.35,3.1){\line(1,0){0.3}}
\put(1.35,3.1){\vector(0,1){1}}
\put(1.35,4.9){\line(1,0){0.5}}
\put(1.35,4.1){\line(0,1){0.8}}

\put(6.65,4.9){\line(-1,0){0.5}}
\put(6.65,4.9){\vector(0,-1){1.0}}
\put(6.65,4.1){\line(0,-1){1.0}}
\put(6.65,3.1){\line(-1,0){0.5}}

\end{picture}

\vspace{-15mm}
\caption{Two compartment model consisting of a lung compartment (gas exchange) and a body compartment with production and metabolism.
Dashed lines indicate equilibrium according to Henry's law.}\label{fig:model_struct}
\end{figure}

To  derive the classical  Farhi equation which relates alveolar concentrations of VOCs to their underlying blood concentrations
one uses a simple two compartment model (see Figure~\ref{fig:model_struct}) which consists of  one single lung compartment and one single body compartment.

\vspace{5mm}

The amount of a VOC transported at time $t$ to and from the lung via blood flow is given by
\begin{eqnarray}
\qc(t)\big(\cven(t)-\cart(t)\big), \nonumber
\end{eqnarray}
where $\qc$ denotes the cardiac output,  $\cven$  is the averaged mixed venous concentration,
  and $\cart$ is the arterial concentration.
  
 On the other hand, the amount exhaled equals
\begin{eqnarray}
\qalv(t)\big(\cinh-\calv(t)\big), \nonumber
\end{eqnarray}
where $\qalv$ denotes the ventilation,  $\cinh$ denotes the concentration in the inhaled air (normally assumed to be zero), and $\calv$ the alveolar air  concentration.

\begin{figure}[ht]
\centering
\begin{picture}(11,2)

\put(9.5,1.5){air}
\put(9.5,0.1){blood}

\put(1,2){\line(1,0){8}}
\put(1,1.3){\line(1,0){2}}\put(7,1.3){\line(1,0){2}}
\put(3,1.3){\line(0,-1){0.6}}
\put(7,1.3){\line(0,-1){0.6}}
\put(1,0.7){\line(1,0){2}}\put(7,0.7){\line(1,0){2}}
\put(1,0){\line(1,0){8}}

\multiput(3.05,1)(0.2,0){20}{\line(1,0){0.1}}

\put(1.2,1.5){$\cinh$}
\put(1.2,0.2){$\cven$}

\put(8.5,1.5){$\calv$}
\put(8.5,0.2){$\cart$}

\put(1.8,1.6){\vector(1,0){6.5}}
\put(5,1.7){$\scriptstyle\qalv$}
\put(1.8,0.3){\vector(1,0){6.5}}
\put(5,0.4){$\scriptstyle\qc$}
\end{picture}
\caption{Diagram of gas exchange in an alveoli symbolized by a dashed line.}\label{fig:gasexchange}
\end{figure}
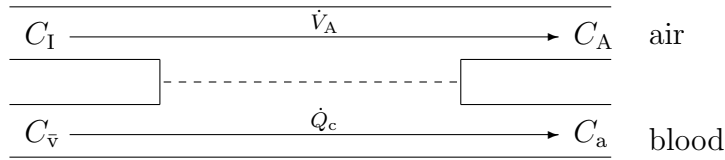

This leads to the following mass balance equation describing the change in the concentration of a VOC in the lung\footnote{For notational convenience we have dropped the time variable  $t$, i.e., we  write $C_{X}$ instead of $C_{X}(t)$, etc. $C_{X}$ denotes the instant or averaged concentration of $X$ over a small sampling period 
$\tau$, i.e.,  
 $C_{X}(t) = 1/ \tau \, \int_{t-\tau/2}^{t+\tau/2} C_{X}(s) ds$.} (see Figure~\ref{fig:gasexchange})
\begin{eqnarray}  \label{eq:alv}
\valv\frac{d\calv}{dt}=\qalv(\cinh-\calv)+\qc(\cven-\cart),
\end{eqnarray}
where $\valv$ denotes the 
volume of the lung.


If the system is in an equilibrium state (e.g., stationary at rest) Equation~(\ref{eq:alv}) reads
$0 = \qalv\big(\cinh-\calv(\cinh)\big)+\qc\big(\cven(\cinh)-\cart\big)$ and using Henry's law $\cart= \hen \calv$ 
we obtain
\begin{eqnarray}
\calv(\cinh) = \frac{  \cinh }{\frac{\hen}{r} + 1}+ \frac{  \cven(\cinh)}{\hen + r}, \label{FarhiCinh}
\end{eqnarray}
where $r=\dot{V}_{A}/\dot{Q}_{c}$ is the ventilation-perfusion ratio and $\hen$ denotes the blood:air partition coefficient. The fact is stressed here that 
$\calv$ and $\cven$ depend on the inhaled concentration $\cinh$. In particular, this means that if $\cinh \not =0$, then subtracting $\cinh$ from $\calv$ to arrive at an estimate for $\calv(0)$ will generally give misleading results (more in Subsection~\ref{subtract}).  

Assuming that $\cinh=0$ we derive the classical Farhi equation  \cite{Farhi1967e}
\begin{eqnarray}
\calv(0) = \frac{\cven(0)}{\hen + r}.
\end{eqnarray}

 We summarize the {\bf assumptions} for the validity of Farhi's equation and the following extensions:
\begin{framed}
\begin{enumerate}
\item the inhaled concentration is zero, i.e., $C_{I}=0$
\item a stationary state is achieved within the lung, i.e., $\frac{ d C_{A} }{d t}  =0$
\item the lung behaves uniformly with respect to ventilation and perfusion (a condition that is typically violated in most lung diseases) 
\item absorption/desorption phenomena within the upper airways are negligible (i.e., low solubility of the VOC in the airway mucus layer, which is generally fulfilled if $\hen<10$, see  \cite{anderson2003})
\item only alveolar air  is sampled so that the alveolar concentration is equal to the exhaled concentration, $\calv= C_{\mathrm{exhaled}}$; in particular, this implies that dead space air contributions have to be avoided, e.g., by CO$_2$ controlled sampling, and that no airway production (as in the case of NO) takes place
\item no reactions with other breath constituents occur, i.e., the VOC under scrutiny is largely inert 
\item the distribution of the blood flow  into the different body compartments remains unchanged (e.g., constant at rest)
\end{enumerate}
\end{framed}

Note that despite its simplicity, the Farhi equation yields first valuable insights into the exhalation kinetics of VOCs. For instance, the breath concentration of compounds with a low blood:gas partition coefficient $\hen$ is expected to react very sensitively to changes of  the  ventilation-perfusion ratio $r$ (e.g., during exercise, hyperventilation, or breath holding \cite{miekisch2014}). Typical examples include methane or butane \cite{methane, King2010GC}.

\subsection{Extension of  the Farhi equation} \label{farhiext}
To calculate the explicit  dependence of $\calv$ and $\cven$ on $\cinh$ we need to consider the mass balance for the body compartment too.
The change of the amount of a VOC in the body is given by the amount which enters the body compartment with the arterial blood plus the amount which is produced in the body minus the amount which is metabolized and the amount leaving via venous blood. Thus the change of the amount of a VOC in the body compartment is given 
by\footnote{Here we used the usual convention to multiply $k_{met}$ by $ \lambda_{b:B}$. It would be more natural to use $k_{met}$ only but this can be incorporated in a redefinition of $k_{met}$.}$^{,}$\footnote{Since the considered inhaled concentrations are low, linear elimination kinetics are sufficient for the description.}
\begin{eqnarray}
\tilde V_{B} \frac{ d C_{B} }{d t} &=  \dot{Q}_{c} (  C_{a}-C_{\bar v}) -\lambda_{b:B} k_{met}  C_{B}+ k_{prod} \label{eq1b},
\end{eqnarray}   
where $k_{met}$ denotes the metabolic rate, $k_{prod}$ the production rate, 
$\tilde V_{B}$ the effective  volume of  the body\footnote{The body blood compartment and the body tissue compartment are assumed to be in an equilibrium and therefore can  be combined into one single body compartment with an effective volume.  For more details about effective volume compare appendix 2 in \cite{king2010a}.},   and $C_{B}$ the concentration 
in the body which is connected to the venous concentration by Henry's law $C_{\bar v} = \lambda_{b:B} \, C_{B}$.
Here $\lambda_{b:B}$  denotes the blood:body tissue partition coefficient.

When in an equilibrium state (i.e., $ \frac{d\calv}{dt}=0$ and $ \frac{dC_{B}}{dt}=0$) we can use  Equations~(\ref{eq:alv}),    (\ref{eq1b}), and $C_{\bar v} = \lambda_{b:B} \, C_{B}$  to eliminate the implicit dependence of $\calv$  on $C_{I}$ in
Equation~(\ref{FarhiCinh}) 

\begin{eqnarray}
C_{A} (C_{I}) & 
= \frac{\frac{k_{prod}}{k_{met}} }{r +\frac{\dot{V}_{A}}{k_{met}} +\hen}+
\frac{r +\frac{\dot{V}_{A}}{k_{met}}}{r +\frac{\dot{V}_{A}}{k_{met}} +\hen} \, C_{I}, \label{eqn5}\\
 C_{\bar{v}} (C_{I})& 
 = \frac{\frac{k_{prod}}{k_{met}}(r + \hen) }{r +\frac{\dot{V}_{A}}{k_{met}} +\hen   } +
\frac{  \frac{\hen}{k_{met}} \, r }{r +\frac{\dot{V}_{A}}{k_{met}} +\hen   } \, C_{I} \label{eqn9}.  
\end{eqnarray}
From Equation~(\ref{eqn5}) and (\ref{eqn9})   we see that the exhaled concentration $C_{A} $ and the mixed venous concentration $ C_{\bar{v}}  $
solely depend on the inhaled concentration $C_{I}$ and the physiological parameters $k_{prod}$, $k_{met}$, $\dot{V}_{A}$, $\dot{Q}_{c}$, $\hen$.

We now discuss some {\em special cases:}
\begin{enumerate}
\item[(a)] For $ C_{I}=0$ (no trace gas is inspired) this reduces to
\begin{eqnarray}
C_{A}(0)
& = \frac{\frac{k_{prod}}{k_{met}} }{r +\frac{\dot{V}_{A}}{k_{met}} +\hen}, \nonumber \\ 
 C_{\bar{v}}(0)  &= \frac{\frac{k_{prod}}{k_{met}} \, (r + \hen) }{r +\frac{\dot{V}_{A}}{k_{met}} +\hen  }
= C_{A}(0)\, (r +\hen)  .
\end{eqnarray}
\item[(b)] On the other hand, when the production is zero ($k_{prod}=0$), this yields
\begin{eqnarray}
C_{A}(C_{I})& 
= \frac{r +\frac{\dot{V}_{A}}{k_{met}}}{r +\frac{\dot{V}_{A}}{k_{met}} +\hen} \, C_{I}
=  \frac{1}{1 +\frac{\hen}{r +\frac{\dot{V}_{A}}{k_{met}}}} \, C_{I}, 
\quad C_{A}(C_{I}) \leq C_{I}, \label{8}\\
 C_{\bar{v}}(C_{I}) & =
\frac{  \frac{\hen}{k_{met}}\, r }{r +\frac{\dot{V}_{A}}{k_{met}} +\hen  } \, C_{I}   =
   \frac{\hen}{k_{met}+\dot{Q}_c}   \, C_{A}(C_{I}) .
\end{eqnarray}
\item[(c)] Assuming $\calv= \cinh$ (zero alveolar gradient) in Equation~(\ref{eqn5}) yields
 \begin{eqnarray}
C_{I}&  =      \frac{k_{prod}}{k_{met}\, \hen}   .
\end{eqnarray}

\end{enumerate}

\subsection{Is subtracing $C_I$ a suitable correction method in order to account for inhaled VOC concentrations?} \label{subtract} 

The contribution of room air concentrations to breath concentrations is a long lasting problem in breath gas analysis (see, e.g., \cite{phillips1999}, \cite{schubert2005}, \cite{beauchamp2014} and the reviews \cite{turner2014}, \cite{pereira2014}). In \cite{phillips1999}, M.~Phillips  summarized  the situation as follows:

{\em Researchers have responded to the problem of room air concentrations with three different strategies:
\begin{enumerate}
\item[(1)] Ignore the problem.
\item[(2)] Provide the subject with VOC-free air to breathe prior to collection of the breath sample. Unfortunately high quality {\em pure} breathing air from commercial sources is usually found to contain a large number of VOCs. In addition it will also contribute to the wash-in/wash-out effect.
\item[(3)] Correct for the problem by subtracting the background VOCs in room air from the VOCs observed in the breath. 
\end{enumerate}}
He calls this difference of exhaled  concentration and inhaled concentration the {\em alveolar gradient},
 i.e., it is assumed that $\calv(0)= \calv(\cinh)-\cinh$. 
To see if this subtraction is correct we consider Equation~(\ref{eqn5}), which we rewrite as 
\begin{eqnarray}
C_{A} (C_{I}) & 
= C_{A} (0)+
\frac{1}{1 +\frac{\hen}{r +\frac{\dot{V}_{A}}{k_{met}}}} \, C_{I}. \nonumber 
\end{eqnarray}
Hence
\begin{eqnarray}
C_{A} (0) &= C_{A} (C_{I}) -\frac{1}{1 +\frac{\hen}{r +\frac{\dot{V}_{A}}{k_{met}}}} \, C_{I}. 
\end{eqnarray}
From this result we conclude that simply {\em subtracting or ignoring} the inhaled concentration is generally false.
More precisely, for VOCs which fulfill the assumptions made above,  $C_{I}$ needs to be multiplied by the following factor
\begin{eqnarray}
a:= \frac{1}{1 +\frac{\hen}{r +\frac{\dot{V}_{A}}{k_{met}}}} 
\end{eqnarray}
before subtraction.
This factor $a$ is approximately $1$ for small values of $\hen$ (e.g., methane, for which $\hen <0.1$) or for small values of $k_{met}$ (no metabolism).
But it might be $2/3$ if, e.g., $\frac{\hen}{r +\frac{\dot{V}_{A}}{k_{met}}}=1/2$.\\
For perspective, Spanel et al.\  experimentally determined $a=0.67$ for isoprene and $a=0.81$ for pentane  \cite{spanel2013}.
Thus one should use the correction $ C_{A} (0)  = C_{A} (C_{I}) -0.67 \, C_{I} $ for isoprene and $ C_{A} (0)  = C_{A} (C_{I}) -0.81 \, C_{I} $ for pentane.

\subsection{Endogenous production and metabolic rates} \label{rates}

The question remains how to determine the endogenous production rate and the total metabolic rate of the body using the theoretical framework introduced above.
When in a stationary state the averaged values of ventilation and perfusion are constant. Thus 
 Equation~(\ref{eqn5})  resembles a straight line of the form

  \begin{eqnarray}
C_{A} (C_{I}) &  =  a \, C_{I} + b, \label{eq17}
 \end{eqnarray}
 $C_{I}$ being the variable here.
The constants $a$ and $b$ are given by
\begin{eqnarray} \label{e5}
&b= C_{A}(0) = \frac{\frac{k_{prod}}{k_{met}} }{r +\frac{\dot{V}_{A}}{k_{met}} +\hen}
 \end{eqnarray}
and
\begin{eqnarray} 
& a = \frac{(r +\frac{\dot{V}_{A}}{k_{met}})}{r +\frac{\dot{V}_{A}}{k_{met}} +\hen}. \label{14a}
 \end{eqnarray}
Thus the constants $a$ and $b$  are completely  determined by the physiological  
quantities $\dot{V}_{A}, \dot{Q}_{c},
k_{prod}, k_{met}$, and $ \hen $. The gradient $a$ is independent of $k_{prod}$,  fulfills $0<a<1$, and is determined by
the metabolic rate $k_{met}$, the ventilation, and  perfusion.
 The quantity $b= C_{A}(0)$ is proportional to the production rate $k_{prod}$.

Varying $C_I$, one can measure $C_{A}(C_I)$   experimentally and thus determine $a$ and $b$.
Measuring in addition ventilation and  perfusion 
 allows for calculating the total production rate and the total metabolic rate of the body from these two equations 
\begin{eqnarray} 
  k_{prod} &= \frac{\frac{b}{1-a}\, \hen \,\dot{V}_{A}}{\frac{a}{1-a  } \,  \hen- r  },\\
  k_{met} &= \frac{\dot{V}_{A}}{\frac{a}{1-a  } \,  \hen- r  }, \label{kmet}
  \end{eqnarray}
  or
\begin{eqnarray}    
  k_{prod} &= (\dot{V}_{A}+(r+\hen)\,  k_{met})\, C_{A}(0), \label{kprod}
 \end{eqnarray}
if $k_{met}$ is known.

Remark 1: In  \cite{spanel2013}, Spanel et al.\ studied the effect of inhaled VOCs on exhaled breath concentrations. 
Unfortunately, breath frequency and heart rate were not reported. Therefore ventilation and perfusion are unknown and thus $k_{prod}$ and $k_{met}$ cannot be estimated. However, this study shows that Equation~(\ref{eqn5}) explains the experimental findings very well.

Remark 2: This approach yields  total endogenous production rates only. As such, one will not be able to determine different production rates in different body compartments.
If more than one production source  exists, a multi compartment model needs to be set up for the body.  Then changes of $r$, e.g., by exercise
will vary the fractional blood flows into these compartments, which subsequently allows for estimating compartment-specific production rates.

{ Remark 3:} Due to the term $(-r)$ in the denominator of Equation~(\ref{kmet}), errors when measuring $a,  \dot{V}_{A}$, and $\dot{Q}_{c}$ may cause  considerable  errors in the rate estimation.

\subsection{Changes in production rates} When measuring breath samples or performing ergometer experiments one assumes that the endogenous  production rate stays constant during the time frame of these experiments. However, when performing breath analysis during sleep it is possible that the production rate will display,  e.g., a circadian rhythm which can be
determined by (ventilation and perfusion are considered to be constant)
  \begin{eqnarray}
 k_{prod} (t) = \big( \dot{V}_{A}  + ( r + \hen)\,  k_{met} \big) \, C_{A,0}(t) . \label{14}
  \end{eqnarray}

\section{Experimental findings} \label{data}

In order to validate the present model, end-tidal concentration profiles of endogenous isoprene and inhaled deuterated isoprene-D5 were obtained by means of a \emph{real-time} setup designed for synchronized measurements of exhaled breath VOCs as well as a number of respiratory and hemodynamic parameters. Our instrumentation has successfully been applied for gathering continuous data streams of these quantities during ergometer challenges 
\cite{King2009} as well as in a sleep laboratory setting \cite{king2012a}. These investigations aimed at evaluating the impact of breathing patterns, cardiac output or blood pressure on the observed breath concentration and at studying characteristic changes in VOCs output following variations in ventilation or perfusion. We refer to~\cite{King2009} for an extensive description of the technical details. 

In brief, the core of the mentioned setup consists of a head mask spirometer system allowing for the standardized extraction of arbitrary exhalation segments, which subsequently are directed into a 
Proton-Transfer-Reaction-Time-of-Flight mass spectrometer
(PTR-MS-TOF, Ionicon Analytik GmbH, Innsbruck, Austria) for online analysis. (The PTR-MS-TOF replaces the formerly used PTR-MS.) This analytical technique has proven to be a sensitive method for the quantification of volatile molecular species $M$ down to the ppb (parts per billion) range by taking advantage of the proton transfer
\[\mathrm{H_3O}^+ + M \to M\mathrm{H}^+ + \mathrm{H_2O}\]
from primary hydronium precursor ions~\cite{lindinger1998,lindinger1998_2}. Note that this ``soft'' chemical ionization scheme is selective to VOCs with proton affinities higher than water (166.5~kcal/mol).
Count rates of the resulting product ions $M\mathrm{H}^+$ or fragments thereof appearing at specified mass-to-charge ratios $m/z$ can subsequently be converted to absolute concentrations of the compound under scrutiny. Specifically, protonated isoprene is detected in PTR-MS-TOF at $m/z=69$,  protonated deuterated isoprene-D5 is detected in PTR-MS-TOF at $m/z=74$ and can be measured with breath-by-breath resolution. 
 An underlying sampling interval of 4~s is set for each parameter. 

For the experiments, deuterated isoprene-D5 (98\%, Campro Scientific GmbH, Germany) was released into the
laboratory room with the help of a 0.5-l glass bulb (Supelco, Canada). In a first step, the bulb was evacuated using a vacuum membrane pump and an
appropriate volume of liquid isoprene (dependent on the target concentration)
was injected through a rubber septum. After complete evaporation of the compound
both Teflon valves of the bulb were opened and the bulb content was purged
with synthetic air at the flow rate of 1~l/min for 3 minutes. Such
conditions provided 3~l of the purge gas (six bulb volumes) to be introduced
into the bulb and, thereby, completely displaced  the original bulb content.
During the bulb purging the laboratory air was continuously mixed with the
help of a fan to achieve a homogenous isoprene distribution.

In contrast to chamber experiments the laboratory serves here as a big reservoir (volume: approx.\ 60 000 l) with a nearly constant background concentration\footnote{For time frames of a few minutes the room air concentration can considered to be constant; however, over one hour a decrease in the room air concentration is noticeable due to leaks in the sealing of the laboratory. }.
Three of the authors (one female, two males) took part in five ergometer sessions each (sessions 1--15), at different room air concentrations of deuterated isoprene-D5 (ranging from 30 to 1000~ppb).
The exact protocol was as follows (see Figure~\ref{fig:isoprened5}): 
 \begin{figure}[h]
\centering
\begin{tabular}{c}
\hspace{-10mm} \includegraphics[height=100mm]{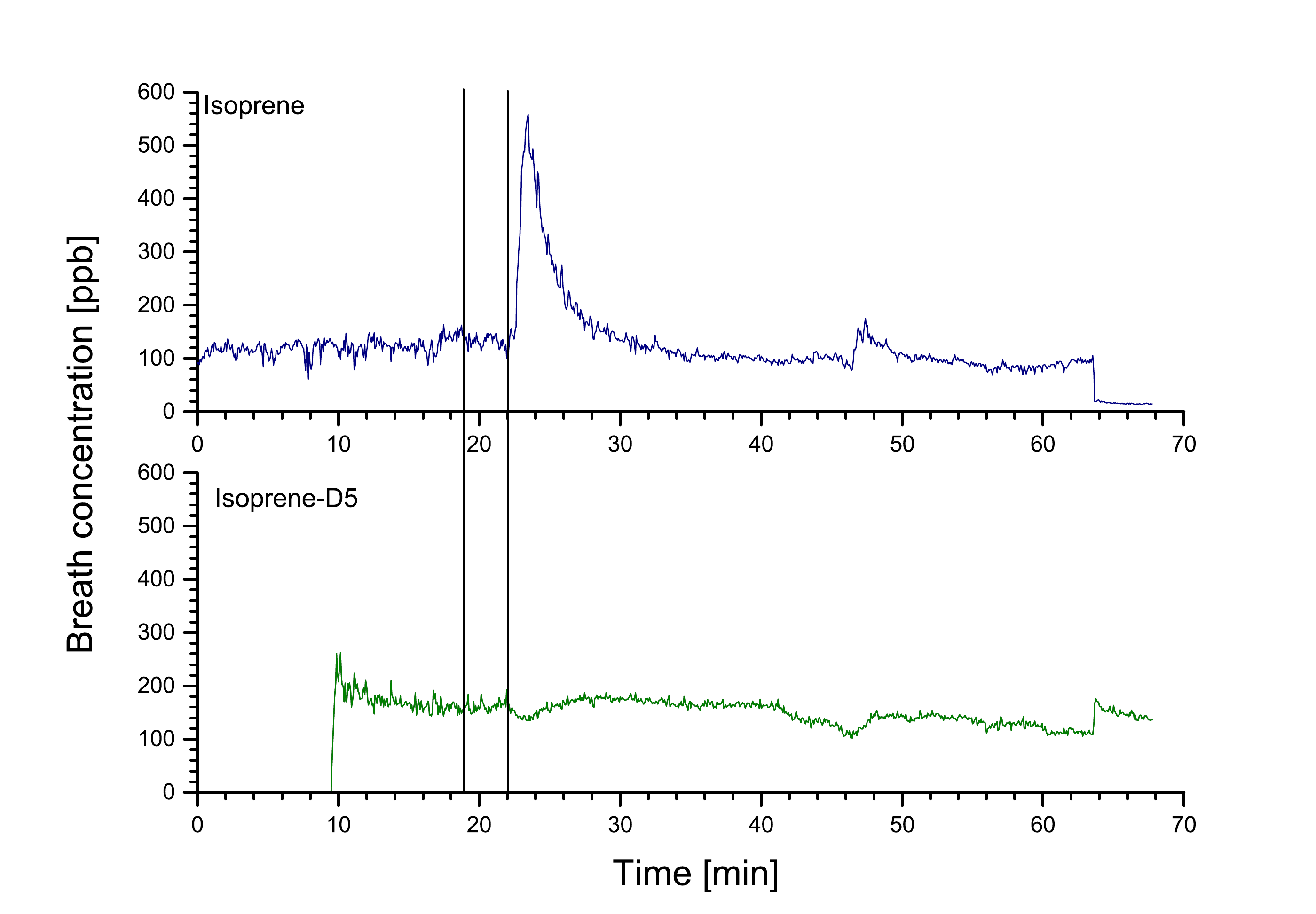}
\end{tabular}
\caption{Typical results of a single ergometer session with inhalation of  deuterated isoprene-D5: rest for 9~min -- release of deuterated isoprene-D5 into the sealed laboratory and waiting for 13~min -- 75~Watts for 18~min -- rest for 6~min -- 75~Watts for 12~min -- rest for 5~min;
deuterated isoprene-D5 breath concentration: green, normal isoprene breath concentration: blue.
In order to validate the 2-compartment model we took the average  values at rest from minute 19 to 22 (vertical lines). These values are given in Tables~\ref{table:volunteer1} -- \ref{table:volunteer3} in column three and four. }\label{fig:isoprened5}
\end{figure} 

\begin{itemize}
\item{} minutes 0--9: the volunteer rests on the ergometer with head-mask on

\item minutes 9--12: deuterated isoprene-D5 is released and the room air is mixed by a fan

\item minutes 12--22: volunteer rests on the ergometer

\item minutes 22--40: volunteer pedals at 75 Watts

\item minutes 40--46: volunteer rests on the ergometer

\item minutes 46--58: volunteer pedals at 75 Watts

\item minutes 58--63: volunteer rests on the ergometer

\item minutes 63--68: mask is taken off and the room air concentration is measured.
\end{itemize}

\section{Results}

As one can deduce from the prototypical plot in Figure \ref{fig:isoprened5}, deuterated isoprene-D5 with a partition coefficient  of nearly $1$ ($\hen=0.95$,  \cite{Mochalski2011a}) enters the arterial blood stream quickly and it takes only a few minutes until it appears in  breath and an equilibrium is achieved in the room air and the blood of the volunteer. To  ensure that a steady state was achieved 
we waited another ten minutes before starting with exercise. At the onset of exercise normal (endogenous) isoprene shows a peak as is well known \cite{King2009}. This peak presumably stems from a high concentration in muscle blood caused by the production in this compartment \cite{King:isoprene, kingmuscle}.
Deuterated isoprene-D5 is nowhere produced in the body. Hence in every compartment of the body its concentration is similar (and zero at the beginning of the experiment).
At the onset of exercise, the ventilation-perfusion ratio goes up and the deuterated isoprene-D5 in exhaled breath 
declines in accordance with the Farhi equation since the venous blood still has an unaltered  isoprene level for 1 to 2 minutes (see minute 22 to 24 in Figure \ref{fig:isoprened5}).
But then, due to the increased inhalation of deuterated isoprene-D5, the venous blood gains a higher concentration level (compare with Equation~(\ref{eqn9})) too 
and the exhaled concentration of deuterated isoprene-D5 reaches its former level (see minute 24 to 40). For perspective, considering that both isoprene compounds can be assumed to have the same blood:gas partition coefficient $\hen$, the profiles in Figure \ref{fig:isoprened5} also show that the exercise peak for normal isoprene cannot be explained by changes in ventilation and perfusion alone. 

The dynamic behaviour in Figure \ref{fig:isoprened5} has mainly been discussed for illustrative purposes. In order to validate the 2-compartment model, only the average resting values of all measured quantities within the last 3 minutes before starting the ergometer challenge were taken into account (minute 19 to 22). These average values are summarized in Tables~\ref{table:volunteer1} -- \ref{table:volunteer3}.

\begin{table}  
\begin{tabular}[t]{|c|c|c|c|c|c|c|}\hline
 &$C_{I, deuterated}$ & $C_{I,normal}$ & $C_{A, deuterated}$ &   $C_{A, normal}$ & $\dot{V}_{A}$ & $\dot{Q}_{c}$\\ 
\hline
&[ppb] & [ppb] & [ppb] & [ppb] & [l/min] & [l/min] \\ \hline \hline
Session 1 &   86.61 & 11.46 &  57.06 & 159.67 & 6.40 & 4.53\\ \hline
Session 2 & 161.88 &  7.01 & 131.47 & 115.12 & 5.31 & 4.37\\ \hline
Session 3 & 202.14 &  5.16 & 156.16 & 100.35 & 9.56 & 6.38 \\ \hline
Session 4 & 447.58 & 8.81 & 288.41 & 137.75  & 8.74 & 4.66 \\ \hline
Session 5 & 935.78 & 12.1 & 390.06 & 114.79 & 6.66 & 4.73\\ \hline
Mean & - & 8.91$\pm$2.93 & - & 125.54$\pm$23.3 & 7.33$\pm$1.76 & 4.93$\pm$0.82 \\ \hline
\end{tabular}
\caption{Volunteer 1 (male, mass: 68~kg, height: 174~cm): normal and deuterated inhaled and exhaled isoprene concentrations with corresponding  ventilation and perfusion values.
}\label{table:volunteer1}
\end{table}
\begin{table}  
\begin{tabular}[t]{|c|c|c|c|c|c|c|}\hline
& $C_{I, deuterated}$ & $C_{I, normal}$ & $C_{A, deuterated}$ &   $C_{A, normal}$ & $\dot{V}_{A}$ & $\dot{Q}_{c}$\\ 
\hline
&[ppb] & [ppb] & [ppb] & [ppb] & [l/min] & [l/min] \\ \hline \hline
Session 6 & 49.81 & 5.48 &  31.77 & 45.53 & 5.75 & 5.69\\ \hline
Session 7 & 104.70 & 6.18 &  71.62 & 50.12 & 5.67 & 6.11\\ \hline
Session 8 & 159.72 & 6.38 & 106.31 & 48.73 & 5.83 & 5.96\\ \hline
Session 9 & 226.08 & 5.74 & 215.86 & 48.76 & 8.95 & 6.88\\ \hline
Session 10 & 515.21 & 7.12 & 213.93 & 36.85 & 8.28 &7.48 \\ \hline
Mean &  - & 6.18$\pm$0.63 & - & 46.0$\pm$5.38 & 6.9$\pm$1.59 &6.42$\pm$0.74 \\ \hline
\end{tabular}
\caption{Volunteer 2 (female, mass: 62~kg, height: 168~cm): 
normal and deuterated inhaled and exhaled isoprene concentrations with corresponding  ventilation and perfusion values.
}\label{table:volunteer2}
\end{table}
\begin{table}
\begin{tabular}[t]{|c|c|c|c|c|c|c|}\hline
& $C_{I, deuterated}$ & $C_{I, normal}$ & $C_{A, deuterated}$ &   $C_{A, normal}$ & $\dot{V}_{A}$ & $\dot{Q}_{c}$\\ 
\hline
& [ppb] & [ppb] & [ppb] & [ppb] & [l/min] & [l/min] \\ \hline \hline
 Session 11 & 32.09 & 7.29 &  22.42 & 184.59 & 8.04 & 4.46\\ \hline
Session 12 &  68.08 & 6.07 &  44.91 & 180.69 & 8.06 & 4.79\\ \hline
Session 13 & 127.22 & 6.37 &  87.93 & 190.25 & 8.65 & 4.54\\ \hline
Session 14 & 164.33 & 5.90 & 137.19 & 142.16 & 7.28 & 4.40\\ \hline
Session 15 & 617.11& 7.81 & 351.69 & 170.88 & 8.13 & 4.09\\ \hline
Mean & - & 6.69$\pm$0.83 & - & 173.71$\pm$19.0 & 8.03$\pm$0.49 & 4.46$\pm$0.25 \\ \hline
\end{tabular}
\caption{Volunteer 3 (male, mass: 90~kg, height: 180~cm): normal and deuterated inhaled and exhaled isoprene concentrations with corresponding  ventilation and perfusion values. 
}
\label{table:volunteer3}
\end{table}

\newpage

From Tables~\ref{table:volunteer1} -- \ref{table:volunteer3} we are able calculate the metabolic rates for  deuterated isoprene-D5 for each volunteer.
To this end, we perform a nonlinear least square optimization using Equation~(\ref{8})
\begin{eqnarray}
\sum \left( \frac{C_{A}(C_{I})}{C_{I}} 
- \frac{(r +\frac{\dot{V}_{A}}{k_{met}})}{r +\frac{\dot{V}_{A}}{k_{met}} +\hen} \right)^2 \longrightarrow \min 
\end{eqnarray}
The sum is taken over the respective sessions for each volunteer here, thereby yielding individual values for $k_{met}$. The results are listed in Table~\ref{table:mrates}.
\begin{table}
\begin{tabular}[t]{|c|c|c|c|}\hline
  & Volunteer 1   &  Volunteer 2 & Volunteer 3   \\ \hline
  $\mbody$    &  $ 16.87\pm 4.8 $&  $7.95\pm 4.0 $& $22.63 \pm 4.8$ \\ \hline
 \end{tabular}
\caption{Metabolic rates (in  [l/min]) for deuterated isoprene-D5 for each volunteer.}\label{table:mrates}
\end{table}
Since normal isoprene and deuterated isoprene-D5 behave similarly from a chemical standpoint, we assume, neglecting isotopic effects, as a first approximation that both have the same metabolic rate.

Using the average resting ventilation $\dot{V}_{A}$, the average resting perfusion $\dot{Q}_{c}$, and the metabolic rates in Table~\ref{table:mrates}, we may thus compute the gradient $ a$ by Equation~(\ref{14a}), and the corrected average exhaled normal isoprene concentration $\calv(0) =\calv(\cinh)- a\, \cinh $. By employing Equation~(\ref{kprod}) we can then calculate the corresponding  endogenous production rate for normal isoprene. The results are listed  in Table~\ref{table:rates}.

\begin{table}
\begin{tabular}[t]{|c|c|c|c|c|c|}\hline
  &  $\dot{Q}_{c}$    &  $\dot{V}_{A}$ &  $  a$ &   $\calv(0)$&$\prbody$\\ 
\hline
           & [l/min] &  [l/min] &  &   ppb&[nmol/min]\\ \hline \hline
Volunteer 1& 4.93 &  7.33& 0.669 &120.6&   216.4\\ \hline
Volunteer 2 & 6.42 & 6.9 & 0.671 &41.9 &   35.7\\ \hline
Volunteer 3 & 4.46 & 8.03 & 0.694 &169.0   &  439.9\\ \hline
\end{tabular}
\caption{Production rates for isoprene for each volunteer (with  $V_{mol}=27$~l).}\label{table:rates}
\end{table}
 
As an additional remark, one can also calculate the total production rate $ k_{prod}$ and the total metabolic rate  $k_{met}$ from the three compartment model presented in\cite{King:isoprene} 
by combining the two body compartments (richly perfused and peripheral compartment)
\begin{eqnarray} \label{eq:combine}
 k_{prod}= \prl+\prm  , \quad k_{met}= \frac{\ml\lrpt\crpt +\mm\lper\cper} {\cven }  \nonumber .
\end{eqnarray} 
Here $\prl, \prm $ denote the production rates in the richly perfused and peripheral compartment, $\lrpt, \lper$ the
corresponding partition coefficients, and $\crpt, \cper$ the corresponding concentrations.

Taking the nominal values from Table \ref{table:volunteer2} and Table C1 in \cite{King:isoprene}  yields $ k_{met}=10$~l/min and $k_{prod}= 125.3$~nmol/min, which is similar to the values extracted above.

\section{Discussion}

In this paper we developed the conceptually simplest compartment model for systemic VOCs that can be described by the Farhi equation in terms of their exhalation kinetics.
In particular, a special focus is given to the case when the inhaled (e.g., ambient air) concentration is significantly different from zero.
The model elucidates a novel approach for computing metabolic/production rates of systemic VOCs with low blood:air partition coefficients
from the respective breath concentrations. Moreover, it clarifies how breath concentration of such VOCS should be corrected when the inhaled concentration cannot be neglected. 
The model  predictions with respect to an affine relationship between exhaled breath concentrations and inhaled concentrations
are in excellent agreement with measurements by Spanel et al.\ \cite{spanel2013}. 

Nevertheless, a number of limitations should be mentioned here. Firstly, in order to apply this model for the estimation of metabolic/production rates,  further studies with a representative number of patients will be necessary. In particular, the individual and population ranges of these quantities will have to be determined.
In addition, it should be investigated how these parameters vary with age, body mass, sex, etc..
To circumvent the intricate measurements of ventilation and perfusion, one could use heart frequency and breath frequency.  

In order to account for long-term exposure, the model should be extended to incorporate a storage compartment which fills up and depletes according to its partition coefficient. 
 This yields then a 3-compartment model. For instance, Pleil et al.\  demonstrated in \cite{Pleil2013b} that a 3-compartment model suffices to model the long-term elimination (over 35 hours) of trichloroethylene after exposure.
However, for short-term exposure experiments as carried out in Section \ref{data}, the influence of such a storage compartment will merely be reflected by a slightly different metabolic rate.
 
 When there is an influence of the upper airway walls (i.e., for highly hydrophilic VOCs), the exhaled concentration deviates 
 considerably from the alveolar concentration, i.e.,  $C_{\mathrm{exhaled}} \not =  C_{A}$.
In that case the lung must be modeled by at least two compartments \cite{king2010a} or more \cite{anderson2003}. In addition breath concentrations will become flow and temperature dependent. Due to this fact, for hydrophilic VOCs one also would have to resort to alternative sampling approaches such as isothermal rebreathing to extract the underlying alveolar concentration \cite{King2010b}. Also, the formulas for metabolic rates and endogenous production rates will be different.

\ack{Acknowledgment}
J.K., P.M.,  M.J., and K.U.\  gratefully acknowledge support from the Austrian Science Fund (FWF) under Grant No.\ P24736-B23.
G.T.\ also gratefully acknowledges support from the Austrian Science Fund (FWF) under Grant No.\ Y330.
We appreciate funding from the Austrian Federal Ministry for
Transport, Innovation, and Technology (BMVIT/BMWA, project 836308, KIRAS). 
We gratefully appreciate funding from the Oncotyrol-project 2.1.1. 
The Competence Centre Oncotyrol is funded within the scope of the
COMET - Competence Centers for Excellent Technologies through BMVIT,
BMWFJ, through the province of Salzburg and the Tiroler
Zukunftsstiftung/Standortagentur Tirol. The COMET Program is conducted
by the Austrian Research Promotion Agency (FFG).
We thank the government of Vorarlberg (Austria) for its generous support.


\appendix
\section{List of symbols}

\begin{table}[H]
\centering 
\begin{tabular}{|lc|}\hline
 {\large\strut} Parameter & Symbol \\ \hline \hline
{\large\strut}	cardiac output & $\qc$   \\
{\large\strut}	averaged mixed venous concentration & $\cven$  \\
{\large\strut}	 arterial concentration & $\cart$ \\
{\large\strut} ventilation & $\qalv$ \\
{\large\strut} inhaled air concentration & $\cinh$ \\
{\large\strut} alveolar air  concentration &  $\calv$ \\
{\large\strut} lung volume &$\valv$ \\
{\large\strut} blood:air partition coefficient &  $\hen$ \\
{\large\strut}  ventilation-perfusion ratio &  $r$ \\
{\large\strut}  exhaled concentration &  $C_{\mathrm{exhaled}}$  \\
{\large\strut} metabolic rate & $k_{met}$   \\
{\large\strut} production rate &  $k_{prod}$ \\
{\large\strut}  effective  volume of  the body\ &  $\tilde V_{B}$ \\
{\large\strut} body concentration &  $C_{B}$  \\
{\large\strut} blood:body partition coefficient &  $\lambda_{b:B} $ \\
{\large\strut} inhaled air concentration of normal isoprene &  $C_{I,normal}$ \\
{\large\strut} inhaled air concentration of  isoprene-D5 &  $C_{I, deuterated}$ \\
{\large\strut} alveolar air concentration of  isoprene-D5 &  $C_{A, deuterated}$ \\
{\large\strut} alveolar air concentration of normal isoprene  &   $C_{A, normal}$ \\
{\large\strut} production rate  in the richly perfused compartment & $ \prl$    \\
{\large\strut} production rate  in the peripheral compartment & $\prm  $ \\
{\large\strut}  blood:richly perfused compartment partition coefficient& $ \lrpt$    \\
{\large\strut} blood:peripheral compartment partition coefficient & $ \lper$   \\
{\large\strut} richly perfused compartment concentration &   $\crpt $     \\
{\large\strut} peripheral compartment concentration &   $ \cper$     \\
\hline
\end{tabular}
\end{table}

\section*{References}
\bibliographystyle{amsplain}
\bibliography{AllCit14}

\end{document}